\documentstyle[sprocl,epsf]{article}
\input{psfig}
\begin{document}
\noindent
{\em Invited talk at the '10th Anniversary HTS Workshop on Physics, Materials
and Applications', Houston, 12-16 march 1996, in press, World Scientific}\\
\title{ELECTRODYNAMICAL PROPERTIES OF HIGH TC SUPERCONDUCTORS STUDIED 
WITH POLARIZED ANGLE RESOLVED INFRARED SPECTROSCOPY}

\author{D. van der Marel, J. Sch\"utzmann, H. S. Somal, and J. W. van der Eb} 

\address{Materials Science Centre, Laboratory of Solid State Physics, \\
University of Groningen, Nijenborgh 4, 9747 AG Groningen, The Netherlands}

\maketitle\abstracts{
Using infrared spectroscopy at grazing angle of incidence we study the
electrodynamical properties of high temperature superconductors. We review some
of our experiments where transverse polarized light is absorbed by
longitudinal optical modes with their mode of oscillation perpendicular
to the plane. This is particularly useful for the study of the plasmons
and phonons perpendicular to the plane, and allows us to study in detail
the c-axis dynamical properties of flux grown single crystals for which 
usually no samples with large dimensions in the c-direction exist.}
\section{Introduction and motivation}
Recently P. W. Anderson pointed out\cite{pwa}, that
for single layer superconductors the following
correlation should exist between the bare Josephson plasmon
energy (in in units of $meV$) and T$_c$ (in $K$) if superconductivity 
is caused by the Anderson-Chakraverty interlayer-tunneling mechanism:

\begin{displaymath}
 \hbar\omega_J = 2.9 \mbox{ } T_c N(0)^{1/2} d^{1/2} a^{-1} 
\end{displaymath} 

\noindent
where $a$ (in $\AA$) is the in-plane lattice parameter, 
$d$ (in $\AA$) is the spacing between
CuO$_2$ planes, and $N(0)$ (in $eV^{-1}$) is the density of states 
at the Fermi energy per unit of CuO$_2$. For all cuprate superconductors 
$N(0)$ is approximately 1 eV, as follows {\em e.g.} from specific heat 
data. Experimentally one observes the
plasma resonance at a reduced value $\omega_J/\sqrt{\epsilon_S}$ due
to screening. In the cuprates this reduction is a factor 3 to 5 depending
on the compound considered. This relation between measurable
quantities is a unique feature of this mechanism, and thus provides 
an experimental test of this theory. 
Using the above expression we constructed the following table:
\begin{displaymath}
\begin{array}{|l|l|l|l|l|l|l|}
\hline
\mbox{Compound} &T_c\mbox{ }(K)&a\mbox{ }(\AA)&d\mbox{ }(\AA)
 &\hbar\omega_J\mbox{ }(meV)&\lambda_c\mbox{ }(\mu m)\\
\hline
\mbox{Bi$_2$Ba$_2$CuO$_6$}    &12 &3.86&11.57&30 &6.6 \\
\mbox{Nd$_{2-x}$Ce$_x$CuO$_4$}&24 &3.95&6.035&43 &4.5 \\
\mbox{La$_{2-x}$Sr$_x$CuO$_4$}&32 &3.79&6.64 &63 &3.1 \\
\mbox{Tl$_2$Ba$_2$CuO$_6$}    &85 &3.86&11.57&216 &0.91 \\
\mbox{Hg$_1$Ba$_2$CuO$_5$}    &98 &3.86&9.51 &225 &0.88 \\
\hline
\end{array}
\end{displaymath}
For optimally doped La$_{2-x}$Sr$_x$CuO$_4$ the value of the {\em un}screened
Josephson plasmon energy is 25-30 meV \cite{tamasaku},
which is not too far below the
prediction based on the Anderson-Chakravarty model.  
For the other systems no direct observations of the Josephson-plasmon energy
have been reported yet, possibly due to the fact that samples of
sufficient thickness for conventional normal incidence reflectivity 
experiments are not available. 
\section{The PARIS method}
In this paper we discuss the reflection properties at grazing angles
of incidence of anisotropic materials. In particular we consider the 
situation where the light is p-polarized, {\em i.e.} with the electric 
field vector parallel to the reflection-plane, and where the dielectric
tensor component of the material along the crystal surface is
metallic-like ($\mbox{Re}\epsilon$ is large and negative).
In this case the absorptivity $A_p$ displays a series
of resonance peaks at frequencies corresponding to the longitudinal 
optical modes with polarization perpendicular to the sample 
surface\cite{comment}. Using the Fresnel equations 
\begin{displaymath}
  \frac{A_p|n_x|\cos{\theta}}{2(2-A_p)} = 
     \frac{\mbox{Im}(\alpha_p)}
     {1+|\frac{\alpha_p}{n_x\cos{\theta}}|^2} 
     \mbox{  }\mbox{  }\mbox{ with }\mbox{  }\mbox{  }
  \alpha_p(\omega) =  e^{i\eta}
  \sqrt{1-\frac{\sin^2\theta}{\epsilon_z}}  
\end{displaymath}
we obtain the pseudo-loss function 
Im$(\alpha_p)/(1+|\alpha_p/n_x\cos{\theta}|^2)$ 
directly from the experimental data, without the need of a Kramers-Kronig
analysis. In this expression $\theta$ is the angle of incidende with the surface normal 
(the z-direction), $\epsilon_i=n_i^2$ is the
dielectric tensor component along $x_i$, and
$\eta \equiv \pi/2- \mbox{Arg}(n_x)$.
If $|\epsilon_z| \gg \sin^2{\theta}$,
this is the loss function Im$(-e^{i\eta}/\epsilon_z)$ with a
Fano-type phase factor. The 
limiting behaviour $\eta=0$ is reached for a superconductor or for
a metal with $\omega \tau \gg 1$. In the low frequency limit
of a metal $\eta=\pi/4$. For a metal with $\theta$ sufficiently far 
below the critical angle we have $|\alpha_p| \ll |n_x|\cos{\theta}$, 
so that the
pseudo-loss function becomes $\mbox{Im} (\alpha_p)$.
Note that for grazing angles of incidence $A_p$ is enhanced 
with a factor $1/\cos\theta$. We recently took 
advantage of this fact to study the in-plane conductivity 
of La$_{2-x}$Sr$_x$CuO$_4$ below T$_c$ in detail.\cite{somal} 
\section{The pseudo-loss function of Tl$_2$Ba$_2$CuO$_6$}
Let us now consider Tl$_2$Ba$_2$CuO$_6$ (T$_c$ = 85 K). 
To simulate the pseudo-loss function at a grazing angle of incidence 
we use the following parameters for the electronic c-axis dielectric function:
$\epsilon_{\infty}=4$,$\hbar\omega_{pc}=200$ (meV),
$\rho_{DC} = 1$ ($\Omega$cm),
and  transverse (longitudinal) optical phonons at 
16.0 (17.7), 51.2 (55.9) and 74.6 (80.3) meV with $\hbar/\tau = 0.6 meV$.   
The result of this simulation is displayed in Fig. 1.
In the normal state the electronic contribution
is overdamped and does not give rise to an additional zero-crossing of 
$\epsilon'$. If we model the superconducting state with BCS theory, as
is shown in the middle curve of Fig. 1,
a Josephson plasmon appears at 6 meV due to transfer of spectral weight
of $\sigma(\omega)$ in the gap-region to the $\delta$-function
at $\omega=0$. The effect of the appearance of an unscreened 
Josephson plasma energy of 200 meV in the superconducting state
is shown in the lower curve: All longitudinal 
modes are now of mixed phonon/plasmon character, and the
longitudinal phonons are pushed to a higher frequency 
compared to the normal state. 
\\
In Fig. 2 the experimentally measured pseudo-loss function
is displayed above and below the phase transition. 
The half-width of the loss-peaks is 
$1/\tau_{ph}+4\pi\sigma_e \epsilon_{\infty}^{-1} 
S_{ph}/(\epsilon_{\infty}+ S_{ph})$,
where $\tau_{ph}$ is the intrinsic phonon life-time, $S_{ph}$ the
oscillator strength, and $\sigma_e$ the electronic optical conductivity.
Hence the line-width of the longitudinal phonons
can be used to obtain an upper-limit
of the electronic contribution to the optical conductivity in the
c-direction. From the experimental data we
obtain an upper-limit for $\sigma_c$ of 1 S/cm near the two
dominant longitudinal phonon-peaks. From comparison with Fig. 1
we conclude that the shift of longitudinal frequencies below 
T$_c$, as well as the occurance of an extra peak at 48 meV,
are both absent in the experiment. This implies that the
unscreened Josephson plasmon energy is below
20 meV, or $\lambda_c > 10 \mu m$ in this material. 
\section{Conclusions} 
The method of polarized angle dependent infrared spectroscopy was used 
to measure the c-axis infrared properties of thin
single-crystalline platelets. The
Josephson plasmon energies were calculated for a number of single
layer compounds adopting an expression suggested by Anderson, and compared
to experimental values for La$_{2-x}$Sr$_x$CuO$_4$ 
and Tl$_2$Ba$_2$CuO$_6$. The agreement
with La$_{2-x}$Sr$_x$CuO$_4$ is within 50 percent. 
For Tl$_2$Ba$_2$CuO$_6$ no shift in longitudinal
phonon frequencies was observed below T$_c$. This may be taken as an
indication, that, as is also the case in conventional BCS theory,
the electronic part of the dielectric function remains almost unaffected
on an energy scale larger than $3.5k_BT_c$.
\section*{Acknowledgments}
We gratefully acknowledge numerous discussions with
P. W. Anderson and stimulating comments by A. J. Leggett, and
N.N. Kolesnikov for supplying Tl$_2$Ba$_2$CuO$_6$ crystals.
This investigation was supported by the Netherlands Foundation for
Fundamental Research on Matter (FOM) with financial aid from
the Nederlandse Organisatie voor Wetenschappelijk Onderzoek (NWO).  

\begin{figure}[t]
 \begin{center}
 \leavevmode
 \hbox{%
  \epsfysize=120mm
  \epsffile{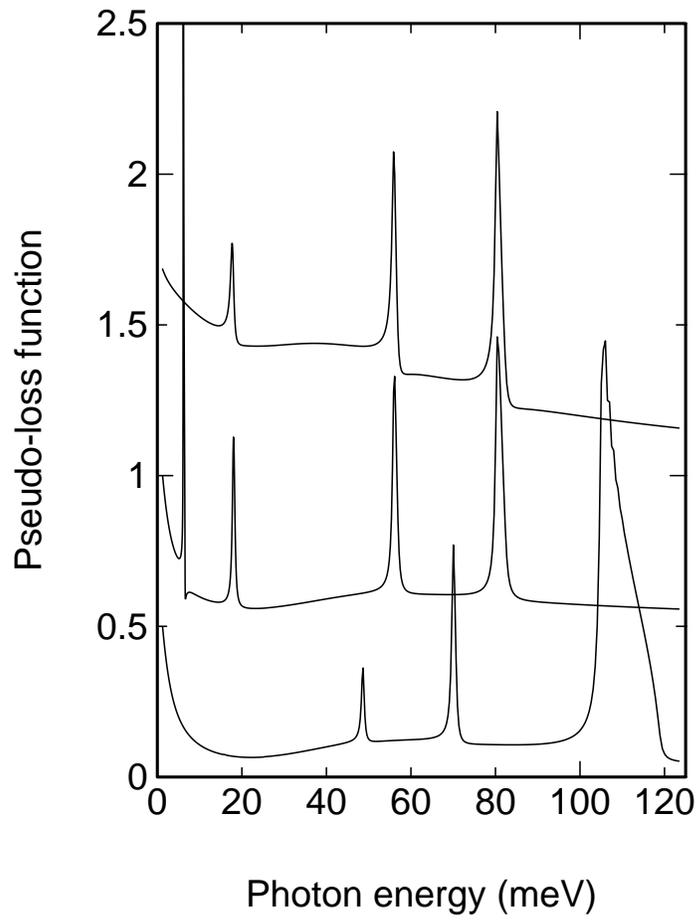}}
 \end{center}
 \caption{Simulation of the pseudo loss-function 
       for Tl$_2$Ba$_2$CuO$_6$. Top curve:
       300 K. Middle curve: 4 K using BCS theory. Lower curve: 4K,
       with $\omega_J=200$ meV.}
\end{figure}
\begin{figure}[t]
  \begin{center}
  \leavevmode
  \hbox{%
    \epsfysize=120mm
    \epsffile{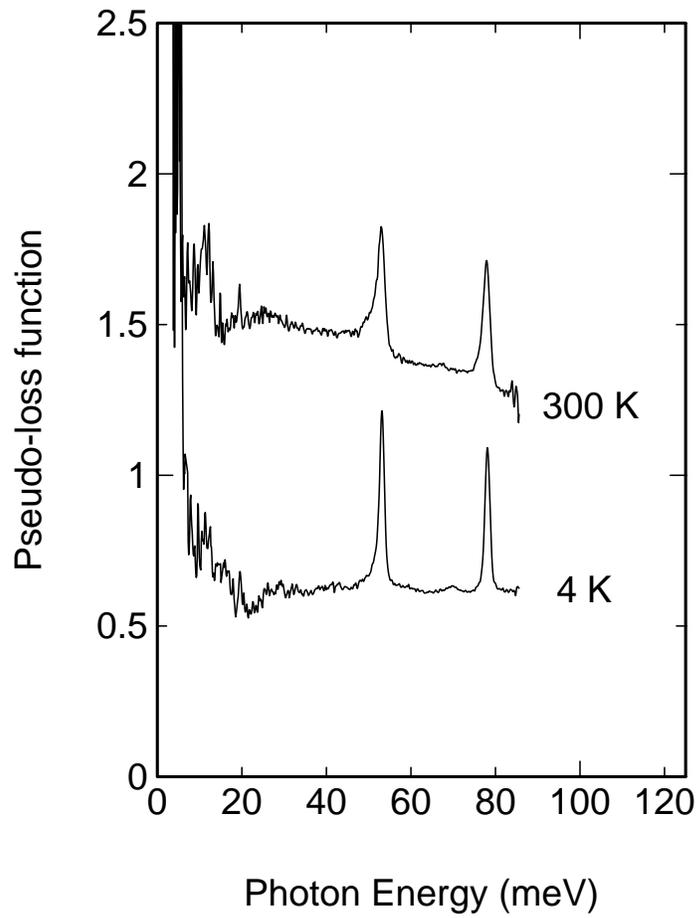}}
  \end{center}
  \caption{Experimental plot of the pseudo-loss function from ab-plane
       surface of Tl$_2$Ba$_2$CuO$_6$ (T$_c$=85 K) 
       using p-polarized light with $\theta = 80^o$.}
\end{figure}
\end{document}